# Spin current control of magnetism

L. Chen[1], Y. Sun[1], S. Mankovsky[2], T. N. G. Meier[1], M. Kronseder[3], H. Ebert[2], D. Weiss[3] and C. H. Back[1,4,5]

[1]*Department of Physics, Technical University of Munich, Munich, Germany*

[2]*Department of Chemistry, Ludwig Maximilian University, Munich, Germany*

[3]*Institute of Experimental and Applied Physics, University of Regensburg, Regensburg, Germany*

[4]*Munich Center for Quantum Science and Technology, Munich, Germany*

[5]*Center for Quantum Engineering, Technical University of Munich, Munich, Germany*

**Exploring novel strategies to manipulate the order parameter of magnetic materials by electrical means is of great importance, not only for advancing our understanding of fundamental magnetism, but also for unlocking potential practical applications. A well-established concept to date uses gate voltages to control magnetic properties, such as saturation magnetization, magnetic anisotropies, coercive field, Curie temperature and Gilbert damping, by modulating the charge carrier population within a capacitor structure[1-5]. Note that the induced carriers are non-spin-polarized, so the control via the electric-field is independent of the direction of the magnetization. Here, we show that the magnetocrystalline anisotropy (MCA) of ultrathin Fe films can be reversibly modified by a spin current generated in Pt by the spin Hall effect. The effect**

**decreases with increasing Fe thickness, indicating that the origin of the modification can be traced back to the interface. Uniquely, the change in MCA due to the spin current depends not only on the polarity of the charge current but also on the direction of magnetization, i.e. the change in MCA has opposite sign when the direction of magnetization is reversed. The control of magnetism by the spin current results from the modified exchange splitting of majority- and minority-spin bands, and differs significantly from the manipulation by gate voltages via a capacitor structure, providing a functionality that was previously unavailable and could be useful in advanced spintronic devices.**

Spin torque (spin transfer torque and spin-orbit torque), which involves the use of angular momentum generated by partially spin-polarized and pure spin-polarized currents, is a well-known method for manipulating the dynamic properties of magnetic materials. In structures such as giant magnetoresistance or tunnel magnetoresistance junctions, the flow of a spin-polarized electric current through the junction imparts spin-transfer torques on the magnetization in the free ferromagnetic layer[6-8]. In heavy metal (HM)/ferromagnet (FM) bilayers, a charge current flowing in the HM induces a spin accumulation at the HM/FM interface, and generates spin-orbit torques acting on the FM[9]. These torques serve as a versatile control mechanism for magnetization dynamics, influencing processes such as magnetization switching[10,11], domain wall motion[12-14], magnetization relaxation (e.g., Gilbert damping)[15], and even auto-oscillations of the magnetization[16,17]. These innovative approaches and their combinations open up a spectrum of possibilities for tailoring magnetic properties with potential implications for advanced technologies such as magnetic random access memories[9,18].

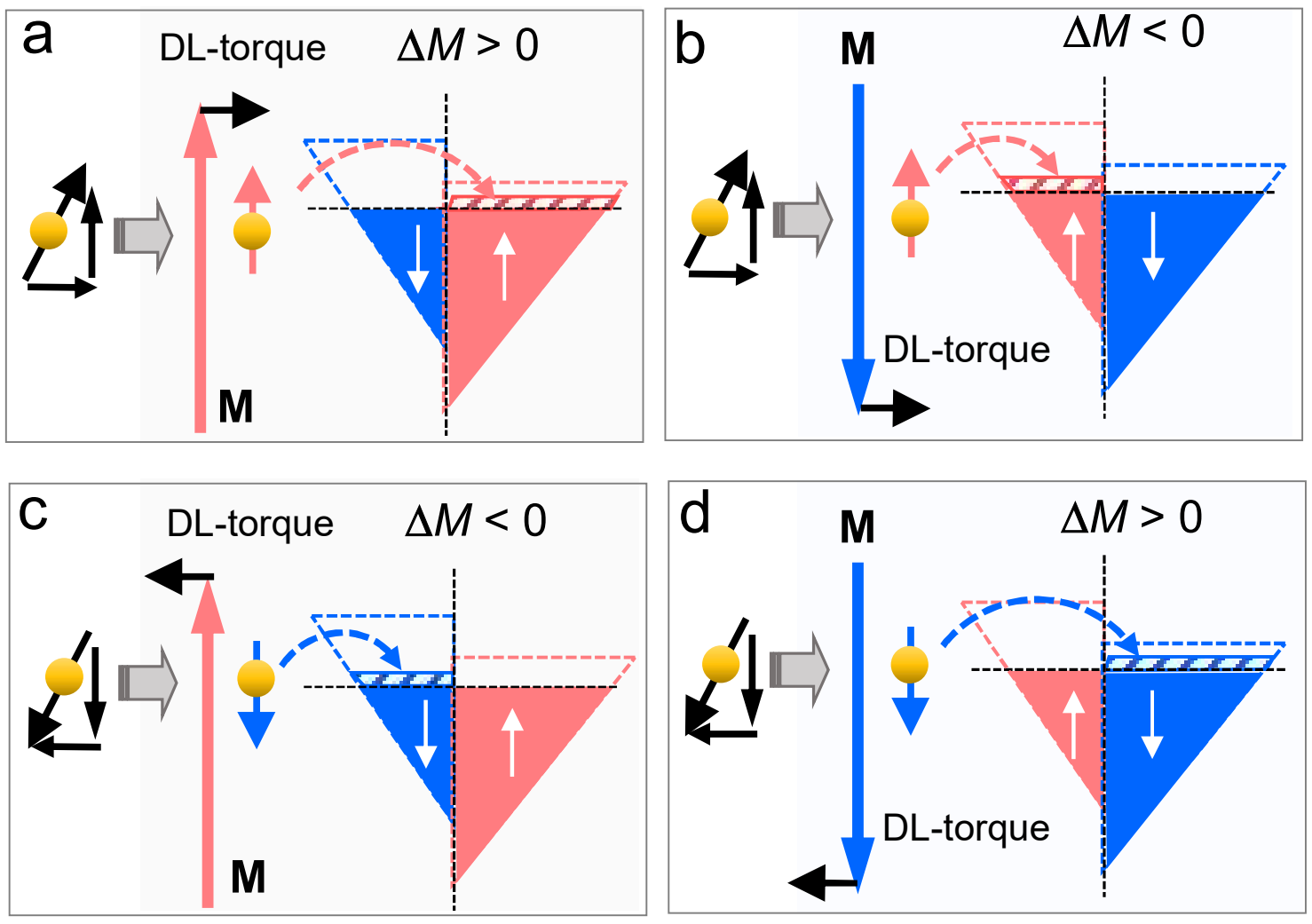

Fig. 1. Schematic of the microscopic mechanism of manipulation and modification of magnetism by a spin current. (**a**) The electron spins transmitted into the FM contain both transverse and longitudinal components with respect to **M**. Due to exchange coupling, the transverse component dephases and is absorbed by **M**, which gives rise to the damping-like spin-orbit torque and is responsible for changing the direction of **M**. The longitudinal component of the spin current is on average aligned with **M**, leading to additional filling of the majority band when **M** is oriented along the +**z**-direction, and an enhancement of the magnitude *M* as well as an increase of magnetic anisotropies are expected due to the enhanced splitting of the majority- and minority-spin energy bands. (**b**) When **M** is aligned along the –**z**-direction, the spin-polarized electron enters the minority band, which can lead to a decrease of *M* as well as a decrease of magnetic anisotropies because of the reduction of the splitting of the majority- and minority-spin energy bands. (**c**) and (**d**): The same as (**a**) and (**b**) but the polarization of the spin current is reversed, which is expected to reduce *M* for **M**//+**z** (c) and to enhance *M* for **M**//-**z** (**d**).

While the impact of a spin current on the orientation of the magnetization (**M**) is widely recognised, there has been no explicit observation so far of successful spin current driven manipulation of the magnitude of **M** – representing the static properties of magnetic materials. To explore this prospect, Fig. 1 illustrates the process of spin current transfer[6-8,19-22]. The spin accumulation generated by a charge current **I**, e.g. by the spin Hall effect (SHE) in the heavy metal, contains both transverse and longitudinal spin components with respect to **M**. The incident transverse spin current dephases and is absorbed by **M**, which gives rise to the damping-like spin-orbit torque and is responsible for the change of the direction of **M**[19,20]. After the spin transfer and within the spin diffusion length of the ferromagnetic metal, the exiting spin current is on average aligned with **M**, and the spin-up electron can fill the majority band when **M** is along the +**z**-direction (Fig. 1a). Due to the enhanced splitting of the majority- and minority-spin bands, this leads to an enhancement of $M$ as well as an increase of the magnetic anisotropies. When **M** is along the −**z**-direction as shown in Fig. 1b, a decrease of $M$ is expected because of the filling of the minority band and the reduction of the splitting of the majority- and minority-spin energy bands. Similarly, once the polarity of the spin current is reversed by reversing the polarity of **I**, a decrease (an increase) of $M$ is expected if **M** // +**z** (−**z**) as shown in Fig. 1c (Fig. 1d). Therefore, the change of the magnetization $\Delta M$ by a spin current is expected to be odd with respect to the inversion of either the charge current or magnetization, i.e., $\Delta M(\mathbf{I}, \mathbf{M}) = -\Delta M(-\mathbf{I}, \mathbf{M}) = -\Delta M(\mathbf{I}, -\mathbf{M})$.

To prove the above scenario, Pt(6 nm)/Al(1.5 nm)/Fe($t_{Fe}$ = 4.5, 2.8, 2.2 and 1.2 nm) multilayers with different Fe thicknesses $t_{Fe}$ are grown on a single two inch semi-insulating GaAs (001) wafer by molecular-beam epitaxy (Fig. 2b and methods). The Pt layer with strong spin-orbit interaction serves as the source of the spin current injected into the Fe layer[22], and is thus responsible for the modification of the dynamic and static magnetic properties. The Al separation layer is used to avoid the magnetic proximity effect between Fe and Pt. Since Al has a weak spin-orbit interaction and a long spin diffusion length, we assume that the spin current injected through the Pt/Al interface is transmitted unchanged to the Al/Fe interface and the transparency for the spin current remains unchanged. The effective mixing conductance $g_{\text{eff}}^{\uparrow\downarrow}$ at the Pt/Al/Fe interfaces, arising form spin pumping, is determined to be $2.7\times10^{18}$ m$^{-2}$, which results in an interfacial spin transparency[23,24] value $T_{\text{int}}$ of 0.21 (Supplementary Note 2).

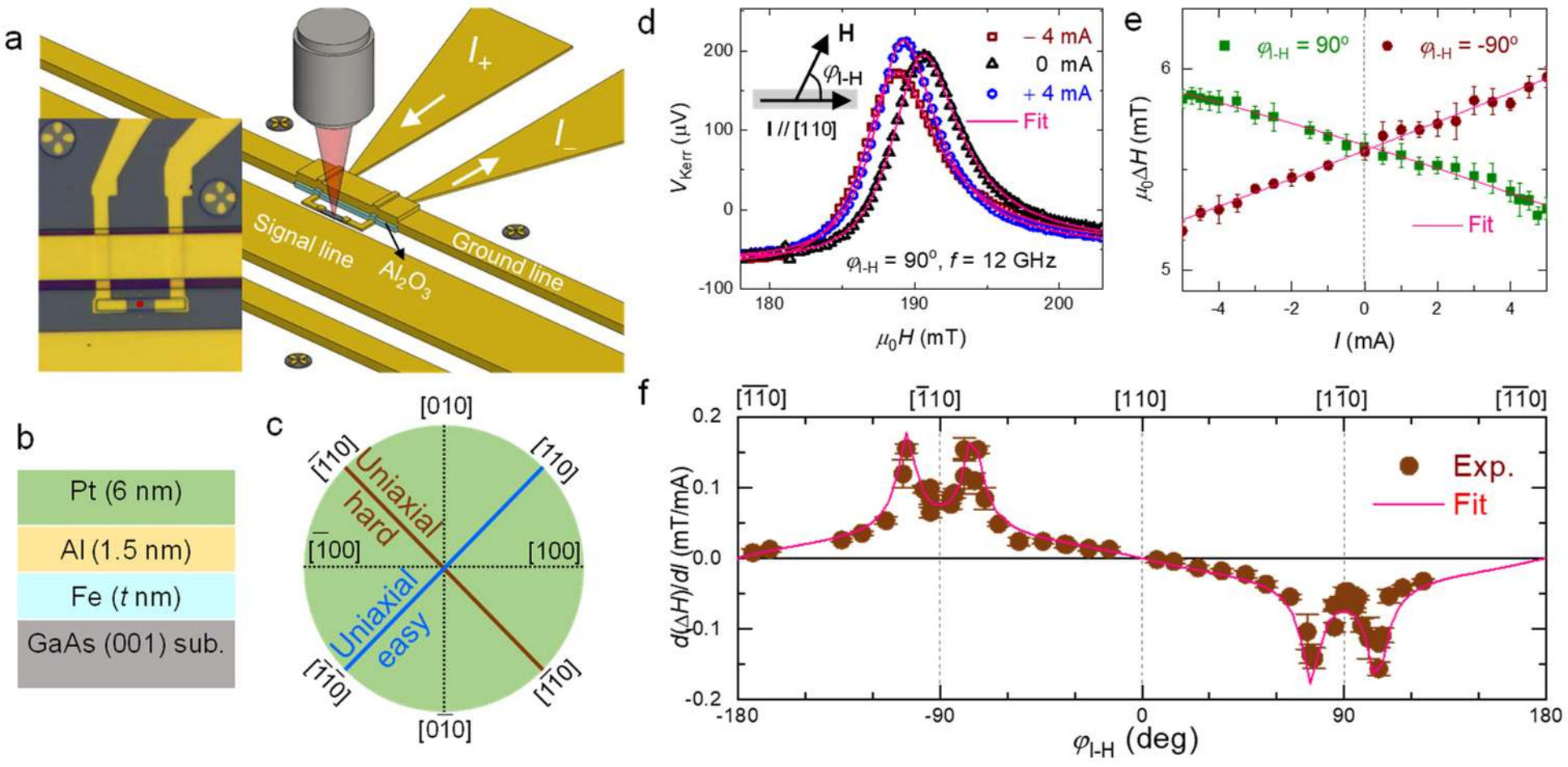

Fig. 2. (**a**) Schematic of the device for the detection of ferromagnetic resonance by time-resolved magneto-optical Kerr microscopy. (**b**) Schematic of the Pt/Al/Fe/GaAs (001) structure. (**c**) Diagram of crystallographic axes with easy and hard magnetization axes along <110> and <$\bar{1}$10> orientations. (**d**) FMR spectra for different dc currents *I* measured at $f$ = 12 GHz and $\varphi_{\text{I-H}}$ = 90°, where $\varphi_{\text{I-H}}$ is the angle between the magnetic-field and the current direction as shown in the inset. (**e**) FMR linewidth (full width at half maximum) as a function of dc current for $\varphi_{\text{I-H}}$ = ±90°; solid lines are the linear fits from which the modulation amplitude $d(\Delta H)/dI$ is obtained. (**f**) $\varphi_{\text{I-H}}$-dependence of $d(\Delta H)/dI$. Error bars show the standard error of the least squares fit. The solid line is the calculation result based on the spin Hall effect of Pt when taking into account the in-plane magnetic anisotropies of Fe (Supplementary Note 4).

The ultra-thin Fe films grown on GaAs (001) substrates allow us to investigate the expected modification of the magnetic properties for two reasons: Fe/GaAs (001) shows I) very low Gilbert damping values $\alpha$ in the sub-nanometer thickness regime ($\alpha$ = 0.0076 for $t_{\text{Fe}}$ = 0.91 nm)[25], and thus it is possible to detect the magnetization dynamics for ultra-thin samples, II) strong interfacial in-plane uniaxial magnetic anisotropy (UMA), which is advantageous for the detection of the spin current induced modification of magnetic anisotropies. The UMA originates from the anisotropic bonding between Fe and As atoms at the GaAs (001) surface[26], where <110>-orientations are the magnetic easy axes (EA) and <$\bar{1}$10>-orientations are the magnetic hard axes (HA) (Fig. 2c). We perform time-resolved magneto-optical Kerr microscopy (TR-MOKE) measurements with out-of-plane driving field to characterize both the static and dynamic magnetic properties of Fe under the influence of spin currents generated by applying a charge current in Pt (Fig. 2a and methods).

Typical ferromagnetic resonance FMR spectra for $t_{\mathrm{Fe}}$ = 2.2 nm and for the charge current applied along a [110]-oriented device are shown in Fig. 2d. The FMR spectra are measured using a fixed microwave excitation frequency $f$ of 12 GHz by sweeping the external magnetic-field $H$ perpendicular to the current, i.e. $\varphi_{\mathrm{I\text{-}H}} = 90^{\circ}$, where $\varphi_{\mathrm{I\text{-}H}}$ is the angle between the current and the magnetic field as defined in the inset of Fig. 2d. A clear modification of the FMR spectrum is observed with the application of a dc current. Each curve is well fitted by combining a symmetric ($L_{\mathrm{sym}} = \Delta H^2 / [4(H-H_{\mathrm{R}})^2 + \Delta H^2]$) and an anti-symmetric Lorentzian ($L_{\mathrm{a\text{-}sym}} = -4\Delta H(H-H_{\mathrm{R}}) / [4(H-H_{\mathrm{R}})^2 + \Delta H^2]$), $V_{\mathrm{Kerr}} = V_{\mathrm{sym}}L_{\mathrm{sym}} + V_{\mathrm{a\text{-}sym}}L_{\mathrm{a\text{-}sym}} + V_{\mathrm{offset}}$, where $H_{\mathrm{R}}$ is the resonance field, $\Delta H$ the full width at half maximum, $V_{\mathrm{offset}}$ the offset voltage, and $V_{\mathrm{sym}}$ ($V_{\mathrm{a\text{-}sym}}$) the magnitude of the symmetric (anti-symmetric) component of $V_{\mathrm{Kerr}}$. It is worth to mention that, by analyzing the position of $H_{\mathrm{R}}$, we have also confirmed that the application of the charge currents does not have detrimental effect on the magnetic properties of the Fe films (Supplementary Note 3).

The dependence of $\Delta H$ on $I$ for $\varphi_{\text{I-H}} = \pm 90^{\text{o}}$ is summarized in Fig. 2e, which shows a linear behavior with opposite slopes for $\varphi_{\text{I-H}} = \pm 90^{\text{o}}$. This reveals the presence of the damping-like spin-orbit torque, confirming previous reports[15]. To extract the modification of the linewidth, the $I$-dependence of $\Delta H$ is fitted by

$$\Delta H = \Delta H_0 + [d(\Delta H)/dI]I + c_1 I^2. \quad (1)$$

Here $\Delta H_0$ is $\Delta H$ for $I = 0$, $d(\Delta H)/dI$ quantifies the modification of linewidth by the spin current and $c_1$ accounts for a possible Joule heating effect on $\Delta H$. A detailed measurement of $d(\Delta H)/dI$ as a function of $\varphi_{\text{I-H}}$ shows that $d(\Delta H)/dI$ varies strongly around the HA. The angular dependence can be well fitted by considering the SHE of Pt with an effective damping-like spin-orbit torque efficiency $\xi$ of 0.06 (Supplementary Note 4). The weaker Bychkov-Rashba-like and Dresselhaus-like spin-orbit interactions arising from the Fe/GaAs interface play a negligible role in the linewidth modulation[27]. Note that the distinctive presence of robust UMA at the Fe/GaAs interface significantly alters the angular dependence of $d(\Delta H)/dI$. This deviation is remarkable when compared to the $\sin\varphi_{\text{I-H}}$-dependence of $d(\Delta H)/dI$ as observed in polycrystalline samples, such as Pt/Py[28,29]. Away from the hard axis (~ $\pm 15^{\text{o}}$), the enhanced sensitivity to SHE results from the fact that the potential barrier in the magnetic energy landscape is lowered upon the application of an external magnetic field (Supplementary Note 4). Consequently, the magnetization behaves 'freely' with no constraints in the vicinity of the HA, since all static torques induced by external magnetic-field and internal magnetic anisotropy fields cancel[30]. In this case, the magnetic stiffness is significantly reduced allowing a large cone angle of precession, which increases the sensitivity to SHE[31].

Having identified the modification of the FMR linewidth by SHE, we now focus on the modification of the resonance field, which is related to the magnetization and magnetic anisotropies of Fe. Figures 3a and b show, respectively, the *I*-dependence of $H_R$ for $t_{Fe}$ = 2.8 nm measured at selected frequencies for *H* applied along the easy axis (**M** // [110]), and along the hard axis (**M** // $[\bar{1}10]$). As shown at the top of each panel column, the current is applied along the [100]-orientation, and the direction of the spin **σ** induced by SHE is along the [010]-orientation with equal projections onto the [110]- and $[\bar{1}10]$-orientations. Therefore, this specific geometry allows a precise comparison of the current-induced modification of the resonance field between [110]- and $[\bar{1}10]$- orientations in the same device. For **M** // [110] as shown in Fig. 3a, all the $H_R$-$I$ traces show a positive curvature; while for **M** // [$\bar{1}$10] in Fig. 3b, traces with a negative curvature are observed. The positive and negative curvatures along the [110] and [$\bar{1}$10]-orientations are due to the fact that the Joule heating induced by the charge current reduces the magnetization and thus the UMA, resulting in an increase of $H_R$ along [110], but a decrease of $H_R$ along [$\bar{1}$10]. Apart from the symmetric parabolic dependence induced by Joule heating, a linear component in the $I$-dependence of $H_R$ is also observed since $H_R(-I) \neq H_R(+I)$ holds. Note that for **M** along both EA and HA, $H_R(-I) > H_R(+I)$ holds for all the measured frequencies. As $t_{Fe}$ is reduced to 1.2 nm, the $I$-dependence of $H_R$ along the EA is similar to the one with $t_{Fe}$ = 2.8 nm and $H_R(-I) > H_R(+I)$ still holds (Fig. 3c). However, for **M** along the [$\bar{1}$10]-orientation as shown in Fig. 3d, the relative magnitude of $H_R(-I)$ and $H_R(+I)$ strongly depends on the excitation frequency, i.e., $H_R(-I) < H_R(+I)$ holds for $f$ = 12.0 GHz; $H_R(-I) \sim H_R(+I)$ holds for $f$ = 14.0 GHz but $H_R(-I) > H_R(+I)$ holds for $f$ = 16.0 GHz. The frequency-dependent shift of the resonance field indicates that the magnetic properties of Fe are modified by the spin current for thinner samples, an observation that has not been reported before.

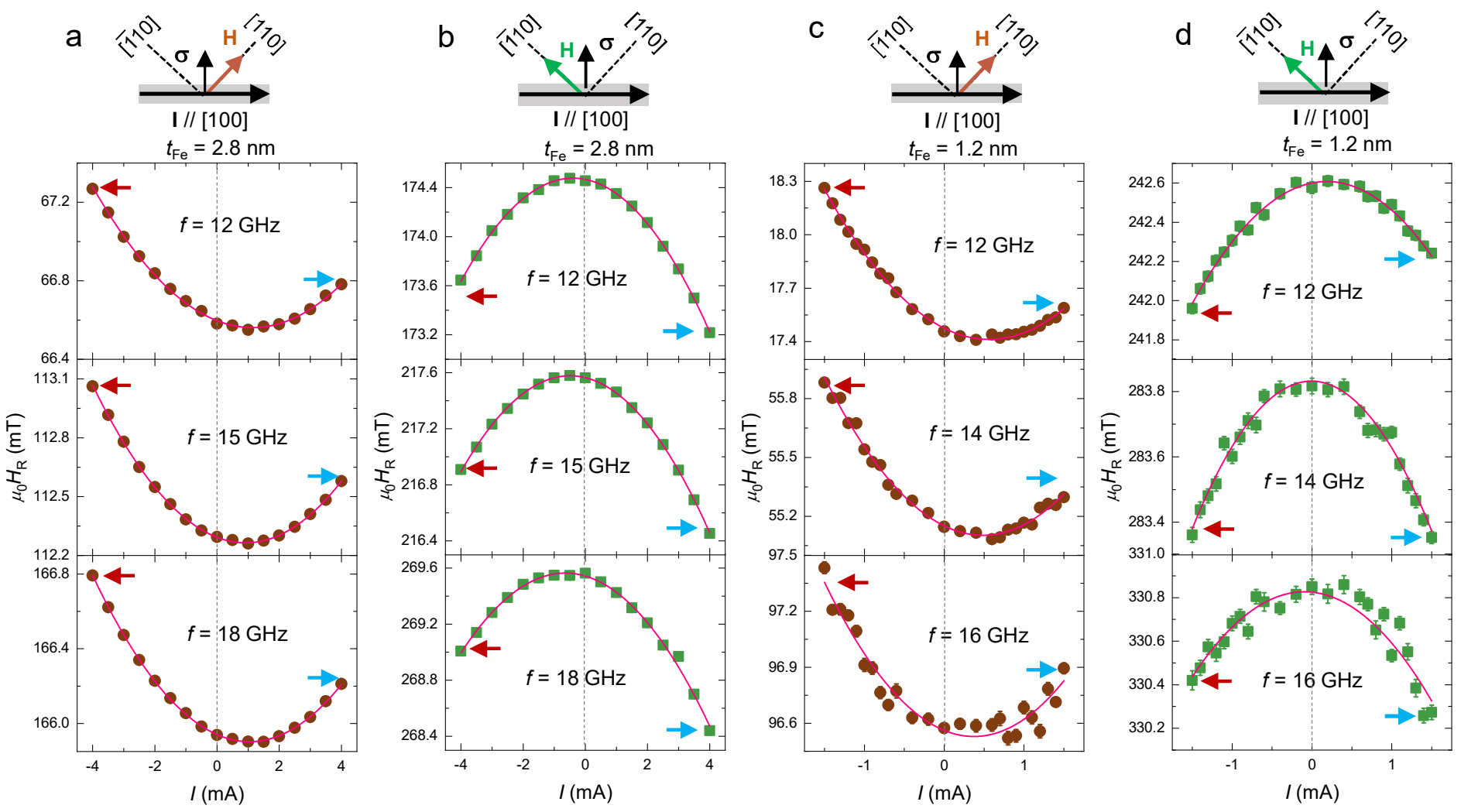

Fig. 3. *I* dependence of $H_R$ measured at selected frequencies for *H* along [110]- (**a**) and $[\bar{1}10]$-orientations (**b**) for $t_{Fe}$ = 2.8 nm. For both field orientations, $H_R(-I) > H_R(+I)$ holds where $H_R(-I)$ and $H_R(+I)$ are respectively marked by red and blue arrows in each panel. (**c**) and (**d**): the same plots as (**a**) and (**b**) but for $t_{Fe}$ = 1.2 nm. In (**c**) for *H* along the [110]-orientation, $H_R(-I) > H_R(+I)$ still holds for all measured frequencies. However, for *H* along the $[\bar{1}10]$-orientation as shown in (**d**), the relative magnitude of $H_R(-I)$ and $H_R(+I)$ depends on the excitation frequency, i.e., for *f* = 12.0 GHz, $H_R(-I) < H_R(+I)$ holds; for *f* = 14.0 GHz, $H_R(-I) \sim H_R(+I)$ holds; while for *f* = 16.0 GHz, $H_R(-I) > H_R(+I)$ holds. As shown in the upper panels, for all the devices, the charge currents are applied along the [100]-orientation, and the direction of the spin accumulation σ is along the [010]-direction with equal projections onto the [110]- and $[\bar{1}10]$-orientations. This experimental trick allows an accurate comparison of the current-induced modification for [110]- and $[\bar{1}10]$-orientations in the same device.

To quantify the modification of the magnetic anisotropies, the $I$-dependence of the $H_R$ trace is fitted by

$$H_R = H_{R0} + (dH_R/dI)I + c_2 I^2. \quad (2)$$

Here, $H_{R0}$ is $H_R$ at $I$ = 0, $dH_R/dI$ quantifies the modification of $H_R$ by the spin current,

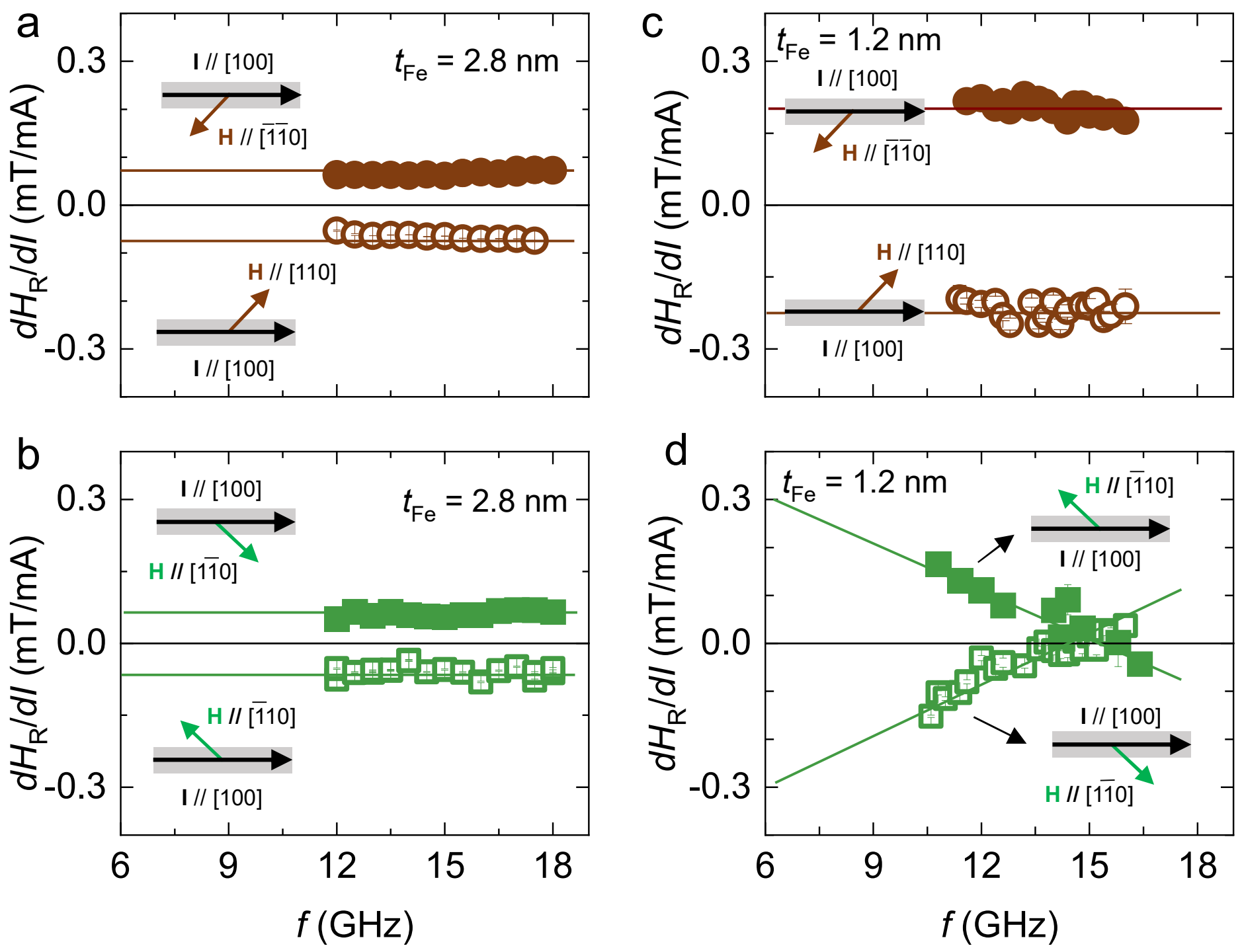

Fig. 4. (**a**) *f*-dependence of $dH_R/dI$ for $H$ along the easy axes ([110]- and [$\bar{1}\bar{1}0$]-orientations). (**b**) *f*-dependence of $dH_R/dI$ for $H$ along the hard axes ([$\bar{1}10$]- and [$1\bar{1}0$]-orientations). The results in (**a**) and (**b**) are obtained for $t_{Fe}$ = 2.8 nm which show that $dH_R/dI$ is independent of $f$ for both easy and hard axes. (**c**) and (**d**) are the same plots as (**a**) and (**b**) but for $t_{Fe}$ = 1.2 nm. The magnitude of $dH_R/dI$ along the hard axes for the thinner sample depends strongly on the excitation frequency, indicating the modification of magnetic anisotropies by spin current. The insets of (**a**)-(**d**) show the relative orientations between the current (**I** // [100], and is represented by black arrows) and the magnetic-field (or magnetization), where the easy axes are represented by brown arrows and the hard axes are represented by green arrows.

and $c_2$ accounts for Joule heating effects on $H_R$. The $f$-dependences of $dH_R/dI$ for different orientations of **M** and different Fe thicknesses are summarised in Fig. 4. For $t_{Fe}$ = 2.8 nm and **M** // <110>-orientations (EA) as shown in Fig. 4a, $dH_R/dI$ is independent of frequency with a positive zero frequency intercept (~ 0.08 mT/mA) for **M** // [$\bar{1}\bar{1}0$]. As **M** is rotated by 180° to the [110]-orientation, the sign of the intercept changes to negative with the same amplitude as the [$\bar{1}\bar{1}0$]-orientation (~ –0.08 mT/mA). This can be understood in terms of the current-induced Oersted field and/or field-like torque $h_{Oe/FL}$, arising from the current flowing in Pt and Al, which shifts $H_R$. The field-like torque originates from the incomplete dephasing (non-transmitted and/or non-dephased) component of the incoming spin[19,20,32]. For **M** along the hard axis as shown in Fig. 4b, the $f$-independent $dH_R/dI$ has also opposite zero-frequency intercepts for **M** // $[\bar{1}10]$ and **M** // $[1\bar{1}0]$ with virtually identical $h_{Oe/FL}$ value as the easy axis. This confirms that the spin accumulation **σ** (along the [010]-orientation) has equal projection onto the $\langle 110\rangle$- and $\langle\bar{1}10\rangle$-orientations. As $t_{Fe}$ is reduced to 1.2 nm (Fig. 4c), the intercept of the $f$-independent $dH_R/dI$ traces along [110] and $[\bar{1}\bar{1}0]$-orientations, respectively, increases to ~ −0.20 mT/mA and ~ 0.20 mT/mA. However, as **M** is aligned along the hard axis as shown in Fig. 4d, the behaviour of the $dH_R/dI$ trace differs significantly from other traces: I) it is no longer $f$-independent but shows a linear dependence on $f$ with opposite slopes for **M** along the $[\bar{1}10]$- and $[1\bar{1}0]$-orientations, II) the absolute value of the zero-frequency intercept along the hard axis (~ 0.32 mT/mA) is no longer equal to that along the easy axis (~ 0.2 mT/mA). The $f$-dependence of the $dH_R/dI$ traces cannot be interpreted to arise from the frequency-independent $h_{Oe/FL}$, and can only be explained by a change of the magnetic anisotropies induced by the spin current.

| Axis | $\Delta H_K$ | $\Delta H_B$ | $\Delta H_U$ |
|---|---|---|---|
| EA | $\Delta H_R = k_K f$ | $\Delta H_R = -\Delta H_B + k_B f$ | $\Delta H_R = \Delta H_U - k_U f$ |
| HA | $\Delta H_R = k_K f$ | $\Delta H_R = -\Delta H_B + k_B f$ | $\Delta H_R = -\Delta H_U$ |

Table 1. Summary of the $\Delta H_R$-$f$ relationships induced by $\Delta H_K$, $\Delta H_U$, and $\Delta H_B$ along easy and hard axes.

In the presence of the in-plane magneto-crystalline anisotropies, the dependencies of $H_R$ on $f$ along the easy axis $H_R^{EA}$ and the hard axis $H_R^{HA}$ are given by the modified Kittel formula[33]

$$\begin{cases} \left(\frac{2\pi f}{\gamma}\right)^2 = \mu_0^2 \left(H_R^{EA} + H_K + \frac{H_B}{2}\right)\left(H_R^{EA} - H_B - H_U\right) \\ \left(\frac{2\pi f}{\gamma}\right)^2 = \mu_0^2 \left(H_R^{HA} + H_K + \frac{H_B}{2} - H_U\right)\left(H_R^{HA} - H_B + H_U\right) \end{cases} \quad (3)$$

where $\gamma$ is the gyromagnetic ratio, $H_K$ the effective magnetic anisotropy field due to the demagnetization field along <001>, $H_B$ the biaxial magnetic anisotropy field along <100>, and $H_U$ the in-plane uniaxial magnetic anisotropy field along <110>. The magnitude of $H_K$, $H_U$ and $H_B$ at $I$ = 0 for each $t_{Fe}$ is quantified by the angular and frequency dependencies of $H_R$ (Supplementary Note 1). Obviously, a change of the magnetic anisotropy fields $H_A$ ($H_A = H_K$, $H_U$, $H_B$) by $\Delta H_A$ ($\Delta H_A = \Delta H_K$, $\Delta H_U$, $\Delta H_B$) leads to a shift of $H_R$ and the magnitude of the shift $\Delta H_R$, defined as $\Delta H_R = H_R(H_A) - H_R(H_A+\Delta H_A)$, depends on the excitation frequency. In the measured frequency range (10 GHz < $f$ < 20 GHz), the $\Delta H_R$–$f$ relations induced by $\Delta H_A$ can be calculated by eq. 3, and their dependencies on $f$ are summarized in Table 1 (Supplementary Note 6). Specifically, an increase (a decrease) of $H_K$ and $H_B$ leads to the $f$-dependent $\Delta H_R$ with positive (negative) slope both along EA and HA. However, a change of $H_U$ leads to the $f$-independent $\Delta H_R$ along HA, while an increase (a decrease) of $H_U$ leads to the $f$-dependent $\Delta H_R$ with negative (positive) slope along EA.

Since $h_{Oe/FL}$, generated by the dc current, also shifts the resonance field along the EA and HA axes by $\pm\frac{\sqrt{2}}{2}h_{Oe/FL}$, where '+' corresponds to the [110] and the [$\bar{1}$10] directions, and '–' corresponds to the [$\bar{1}\bar{1}$0] and the [1$\bar{1}$0] directions, the total $\Delta H_R$ along EA and HA is given by

$$\begin{cases} \Delta H_{\mathrm{R}}^{\mathrm{EA}}(f) = \Delta H_{\mathrm{U}} - \Delta H_{\mathrm{B}} \pm \frac{\sqrt{2}}{2} h_{\mathrm{Oe/FL}} + (k_{\mathrm{K}} + k_{\mathrm{B}} - k_{\mathrm{U}}) f \\ \Delta H_{\mathrm{R}}^{\mathrm{HA}}(f) = -(\Delta H_{\mathrm{U}} + \Delta H_{\mathrm{B}}) \pm \frac{\sqrt{2}}{2} h_{\mathrm{Oe/FL}} + (k_{\mathrm{K}} + k_{\mathrm{B}}) f \end{cases} \tag{4}$$

Here the slope $k$ [$k$ = $k_{\mathrm{K}}$, $k_{\mathrm{U}}$, $k_{\mathrm{B}}$, and $k = \frac{d(\Delta H_R)}{df}$] quantifies the modulation of $H_{\mathrm{R}}$ induced by $\Delta H_{\mathrm{A}}$. For $\Delta H_{\mathrm{R}}$ induced by both $\Delta H_{\mathrm{K}}$ and $\Delta H_{\mathrm{B}}$, $k \propto \Delta H_A$, which holds for both EA and HA. However, for $\Delta H_{\mathrm{R}}$ induced by $\Delta H_{\mathrm{U}}$, we find $k \propto -\Delta H_U$, which holds only for EA. Since the $f$-dependence of $\Delta H_{\mathrm{R}}^{\mathrm{EA}}$ induced by $\Delta H_U$ has an opposite slope as those induced by $\Delta H_{\mathrm{K}}$ and $\Delta H_{\mathrm{B}}$, it is possible to obtain a $f$-independent $\Delta H_{\mathrm{R}}^{\mathrm{EA}}$ along EA (Fig. 4c) by tuning the corresponding parameters and to obtain a $f$-linear $\Delta H_{\mathrm{R}}^{\mathrm{HA}}$ along HA (Fig. 4d). To reproduce the results along the [110]- and [$1\bar{1}0$]-orientations (i.e., the net magnetization of these two orientations is parallel to **I**) for $t_{\mathrm{Fe}}$ = 1.2 nm, we obtain $\Delta H_{\mathrm{B}}$ = 0.26 mT/mA, $\Delta H_{\mathrm{K}}$ = 2.0 mT/mA and $\Delta H_{\mathrm{U}}$ = 2.5 mT/mA through eqs. 3 and 4 (for details see Supplementary Note 6). In contrast, for the data sets for **M** along [$\overline{11}0$] and [$\bar{1}10$]-orientations (i.e., the magnetization is rotated by 180° and the net magnetization of these two orientations is antiparallel to **I**), $\Delta H_{\mathrm{B}}$ = –0.26 mT/mA, $\Delta H_{\mathrm{K}}$ = –2.0 mT/mA and $\Delta H_{\mathrm{U}}$ = –2.5 mT/mA are obtained, which have the opposite polarity compared to the case of **M** along the [110]- and [$1\bar{1}0$]-orientations.

Figure 5 summarizes the obtained modification of the magnetic anisotropy $\Delta H_{\mathrm{A}}$ as a function of $t_{\mathrm{Fe}}$. For $t_{\mathrm{Fe}}$ above 2.8 nm, the modification of the magnetic anisotropy is too small to be observed and is washed out by bulk effects. For $t_{\mathrm{Fe}}$ below 2.2 nm, the modification of $\Delta H_{\mathrm{A}}$ increases as $t_{\mathrm{Fe}}$ decreases. This indicates that the spin current induced modification of the magnetic energy landscape is of interfacial origin, similar to the damping-like spin-torque determined by the $f$-dependence of $d(\Delta H)/dI$ in the same sample series (Supplementary Note 5). The modification changes sign when **M** is rotated by 180°, which fully validates the scenario of $\Delta H_A(\mathbf{I}, \mathbf{M}) = -\Delta H_A(-\mathbf{I}, \mathbf{M}) = -\Delta H_A(\mathbf{I}, -\mathbf{M})$ as suggested in Fig. 1. For a given **M** direction, the obtained $\Delta H_B$, $\Delta H_K$, and $\Delta H_U$ have the same sign, which is also consistent with a monotonic increase (decrease) of $H_B$, $H_K$, and $H_U$ as temperature decreases (increases) (Supplementary Note 6). Moreover, these results also show that $H_U$ is more sensitive to the spin current than $H_K$ and $H_B$, highlighting the importance of UMA to enable the observation of the modifications. The much smaller $\Delta H_B$ value is due to the fact that $H_B$ is 1-2 orders smaller than $H_U$ and $H_K$ in the ultra-thin regime (Supplementary Note 1).

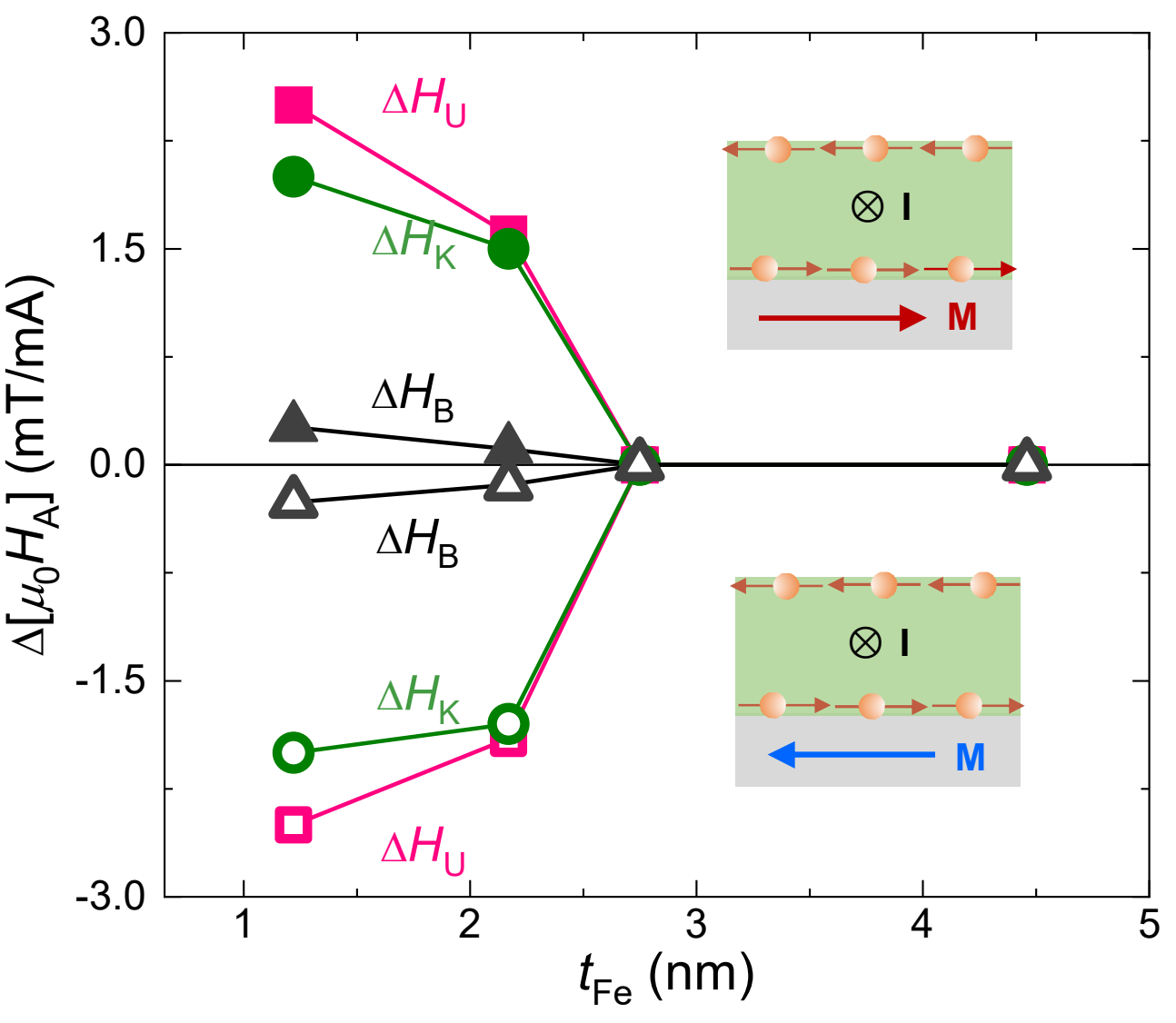

Fig. 5. Summary of $t_{Fe}$ dependence of $\Delta H_A$ ($\Delta H_A = \Delta H_K$, $\Delta H_U$, $\Delta H_B$) for opposite magnetization **M** directions, where solid symbols represent **M** // +**z**-direction and open symbols represent **M** // −**z**-direction. The relative orientations between the charge current **I** and **M** are shown in the inset.

The change of magnetic anisotropy $\Delta H_K$ is directly related to the change of magnetization $\Delta M$, which can be attributed to the additional filling of the electronic *d*-band. Strictly speaking, the induced filling of the bands in Fe occurs mainly close to the interface and is not homogeneously distributed, since it depends on the spin diffusion length of the spin current in Fe. In other words, the measured modulated magnetic anisotropies are averaged over the whole ferromagnetic film. To first-order approximation and for simplicity, we neglect the spin current distribution in Fe and assume that it is homogeneously distributed. The spin accumulation at the interface[9] is given by $u_s^0 = 2e\lambda\xi E \tanh\left[\frac{t_{pt}}{2\lambda}\right]$, where $e$ is the elementary charge, $\lambda$ the spin diffusion length, $E$ (= $j/\sigma$) the electric-field, $j$ the current density and $\sigma$ the conductivity of Pt. The areal spin density $n_{SHE}$ transferred into Fe, is obtained as $n_{SHE} = u_s^0\lambda N$[9], where $N$ is the density of states at the Fermi level for Fe. Using $N = 6 \times 10^{48}$ $J^{-1}m^{-3}$, $\lambda = 4$ nm, $\xi = 0.06$, $\sigma = 2.0 \times 10^6$ $\Omega^{-1}m^{-1}$, $n_{SHE} = 4.2 \times 10^{12}$ $\mu_B/cm^2$ is obtained for $I = 1$ mA. Since Fe has a bcc structure (lattice constant $a = 2.8$ Å) with a moment of ~1.0 $\mu_B$ for $t_{Fe}$ = 1.2 nm at room temperature, the areal density of the magnetic moment of Fe $n_{Fe}$ is determined to be $2.6 \times 10^{14}$ $\mu_B/cm^2$. In this case, the filling of the *d*-band by SHE leads to a change of the magnetic moment of the order of $n_{SHE}/n_{Fe} \sim 0.16\%$, which agrees with the ratio between $\Delta H_K$ and $H_K$, i.e., $\Delta H_K/H_K \sim 2.0$ mT/ 1 T ~ 0.2%.

On the other hand, since the UMA shows a larger modulation than effective demagnetization and biaxial anisotropy, it is clear that the band structure of the interfacial Fe/GaAs plays a major role for the observed effect when the Fe film is sufficiently thin[34]. As shown in Fig. 1a, the spin current $\mathbf{J}_S^z$ can be represented in terms of two opposite flows of electrons carrying up and down spin moments, $\mathbf{J}_S^z = \mathbf{J}_{up} - \mathbf{J}_{dw}$, in the absence of a net electric current ($\mathbf{J}_c = \mathbf{J}_{up} + \mathbf{J}_{dw} = 0$). Since the spin current doesn't transfer charge, the inflow of the spin-up electrons leads to an increase of occupation of the spin-up d-states of Fe, while the outflow of spin-down electrons leads to a decrease of occupation of the spin-down d-states. A similar effect occurs in the presence of an applied magnetic field on the order of the molecular magnetic-field, i.e., **H** // +**z** shifts the spin-up d-states down in energy, following by an increase of their occupation. In contrast, the spin-down d-states shift up in energy, leading to their depopulation. Therefore, to mimic the effect of spin current on the UMA and magnetic moment, we have investigated the dependence of the UMA on the external magnetic-field on the basics of first-principle electronic band structure calculations for Fe/GaAs. The resulting modification of UMA has been determined by means of magnetic torque calculations[35](Supplementary Note 7). The applied magnetic-field results in an increase of the magnetic anisotropy energy along the easy axis, if the applied magnetic-field is parallel to the magnetization direction, and to a decrease of anisotropy in the case of antiparallel orientation. These changes are accompanied by an increase (for $H > 0$) or decrease (for $H < 0$) of the spin magnetic moment of Fe, consistent with experimental observations. Additional more sophisticated model might be needed to extend the existing model and to explain the experimental results quantitatively.

Our results have shown that the intrinsic properties of ultra-thin ferromagnetic materials, i.e., the magnitude of **M** and the magnetocrystalline anisotropies, can be varied in a controlled way by spin currents, which has been ignored in the spin-transfer physics so far. This unique route of controlling magnetic anisotropies is not accessible by other existing ways using electric-field[1-5] and mechanical stress[36,37] in which the control of magnetism is independent on magnetization direction. Besides the magnitude of the magnetization, other material parameters, e.g., the Curie temperature, coercive field etc., are also expected to be controllable by spin current. It is known that spin-torque plays an essential role in modern spintronic devices, beyond this proof of principle, the so far unnoticed modification of the length of the magnetization vector by spin currents could offer an alternative and attractive generic actuation mechanism for spin-torque phenomena, e.g., magnetization switching and auto-oscillations of the magnetization. We expect such a modification of the magnetic energy landscape to be a general feature, not only limited to ferromagnetic metal/heavy metal systems with strong spin-orbit interaction but also present in the case of conventional spin-transfer torques, where it is generally believed that the magnitude of **M** is fixed during the spin transfer process[6-8]. Moreover, the modification is not limited to in-plane ferromagnets, and one could manipulate the static magnetic properties of ferromagnets with perpendicular anisotropy by using out-of-plane polarized spin current source, e.g., $WTe_2$[38], $RuO_2$[39-42] and non-linear antiferromagnets $Mn_3Sn$[43] and $Mn_3Ga$[44]. Finally, we believe that much larger modification amplitudes can be realized in other more effective spin current sources based on the wide-range of spin-torque material choices[9].

## Methods

**Sample preparation.** Samples with various Fe thicknesses $t_{Fe}$ are grown by molecular-beam epitaxy (MBE). First a GaAs buffer layer of 100 nm is grown in a III-V MBE, after that the GaAs substrate (semi-insulating wafer, which has a resistivity $\rho$ between $1.72\times10^8$ Ω.cm and $2.16\times10^8$ Ω.cm) is transferred to a metal MBE without breaking the vacuum for the growth of the metal layers. For a better comparison of physical properties of different samples, various Fe thicknesses are grown on a single two-inch wafer by stepping the main shadow shutter of the metal MBE. After the growth of the step-wedged Fe film, 1.5-nm Al/6-nm Pt layers are deposited on the whole wafer.

**Device.** First, Pt/Al/Fe stripes with a dimension of 4.0 μm × 20 μm and with the long side along the [110]- and [100]-orientations are defined by a mask-free writer and Ar-etching. After that, contact pads for the application of the dc current, which are made from 3-nm Ti and 50 nm Au, are prepared by evaporation and lift-off. Then, a 70-nm $Al_2O_3$ layer is deposited by atomic layer deposition to electrically isolate the dc contacts and the coplanar waveguide (CPW). Finally, the CPW consisting of 5 nm Ti and 150 nm Au is fabricated by evaporation, and the Fe/Al/Pt stripes are located in the gap between the signal line and ground line of the CPW (Fig. 2a). The CPW is designed to match the rf-network which has an impedance of 50 Ω. The width of the signal line and the gap is 50 μm and 30 μm, respectively. Magnetization dynamics of Fe is excited by out-of-plane Oersted field induced by the rf microwave currents flowing in the signal and ground lines.

**Measurements.** For TRMOKE microscopy measurements, a pulse train of a Ti:Sapphire laser (repetition rate of 80 MHz and pulse width of 150 fs) with wavelength of 800 nm is phase-locked to the microwave current. A phase shifter is used to adjust the phase between the laser pulse train and microwave, and the phase is kept constant during the measurement. The polar Kerr signal at a certain phase, $V_{\mathrm{Kerr}}$, is detected by a lock-in amplifier by phase modulating the microwave current at a frequency of 6.6 kHz. The $V_{\mathrm{Kerr}}$ signal is measured by sweeping the external magnetic field, and the magnetic-field can be rotated in-plane by 360°. A Keithley 2400 device is used as the dc current source for linewidth and resonance field modifications. All measurements are performed at room temperature.

## References

1. Ohno, H., Chiba, D., Matsukura, F., Omiya, T., Abe, E., Dietl, T., Ohno, Y., & Ohtani, K., Electric-field control of ferromagnetism, *Nature* **408**, 944 (2000).
2. Chiba, D., Sawicki, M., Nishitani, Y., Nakatani, Y., Matsukura, F., & Ohno, H., Magnetization vector manipulated by electric fields, *Nature* **455**, 515-518 (2008).
3. Weisheit, M. *et al*. Electric field-induced modification of magnetism in thin-film ferromagnets, *Science* **315**, 349 (2007).
4. Chen, L., Matsukura, F., & Ohno, H., Electric-field modulation of damping constant in a ferromagnetic semiconductor (Ga,Mn)As, *Phys. Rev. Lett.* **115**, 057204 (2015).
5. Matsukura, F., Tokura, Y., & Ohno, H., Control of magnetism by electric-fields, *Nature Nanotech*. **10**, 209-220 (2015).
6. Berger, L., Emission of spin waves by a magnetic multilayer traversed by a current, *Phys. Rev. B* **54**, 9353 (1996).
7. Slonczewski, J. C., Current-driven excitation of magnetic multilayers, *J. Mag. Mag. Mater*. **159**, L1-L7 (1996).
8. Ralph, D. C., & Stiles, M. D., Spin transfer torques, *J. Mag. Mag. Mater*. **320**, 1190-1216 (2008).
9. Manchon, A. *et al*. Current-induced spin-orbit torques in ferromagnetic and antiferromagnetic systems, *Rev. Mod. Phys*. **91**, 035004 (2019).
10. Miron, I. M. *et al*. Perpendicular switching of a single ferromagnetic layer induced by in-plane current injection. *Nature* **476**, 189-193 (2011).
11. Liu, L. Q. *et al*. Spin-torque switching with giant spin Hall effect of Tantalum. *Science* **336**, 555-558 (2012).
12. Miron, I. M. *et al*. Fast current-induced domain-wall motion controlled by Rashba effect. *Nature Mater.* **10**, 419-423 (2011).

13. Emori, S. *et al*. Current-driven dynamics of chiral ferromagnetic domain walls. *Nature Mater*. **12**, 611-616 (2013).

14. Ryu, K. *et al*. Chiral spin torque at magnetic domain walls. *Nature Nanotech.* **8**, 527-533 (2013).

15. Ando, K. *et al*. Electric manipulation of spin relaxation using the spin Hall effect, *Phys. Rev. Lett*. **101**, 036601 (2008).

16. Demidov, V. E. *et al*. Magnetic nano-oscillator driven by pure spin current, *Nature Mater*. **11**, 1028-1031 (2012).

17. Liu, L. Q. *et al*. Magnetic Oscillations driven by the spin Hall effect in 3-terminal magnetic tunnel junction devices, *Phys. Rev. Lett.* **109**, 186602 (2011).

18. Shao, Q. *et al*. Roadmap of spin-orbit torques, *IEEE Transactions on magnetics*, **57**, 1-39 (2021).

19. Amin, V. P., & Stiles, M. D., Spin transport at interfaces with spin-orbit coupling: Formalism, *Phys. Rev. B* **94**, 104419 (2016).

20. Amin, V. P., & Stiles, M. D., Spin transport at interfaces with spin-orbit coupling: Phenomenology, *Phys. Rev. B* **94**, 104420 (2016).

21. Haney, P. M. *et al*. Current induced torques and interfacial spin-orbit coupling: Semiclassic modeling. *Phys. Rev. B* **87**, 174411 (2013).

22. Sinova, J., Valenzuela, S. O., Wunderlich, J., Back, C. H., & Jungwirth, T., Spin Hall effects, *Rev. Mod. Phys*. **87**, 1213 (2015).

23. Pai, C. F. *et al*. Dependence of efficiency of spin Hall torque on the transparency of Pt/ferromagnetic layer interfaces. *Phys. Rev. B* **92**, 064426 (2015).

24. Zhu, L., Ralph, D. C., & Buhrman, R. A., Effective spin-mixing conductance of heavy-metal-ferromagnet interfaces, *Phys. Rev. Lett*. **123**, 057203 (2019).

25. Chen, L., Mankovsky, S., Kronseder, M., Schuh, D., Prager, M., Bougeard, D.,

Ebert, H., Weiss, D., and Back, C. H., Interfacial tuning of anisotropic Gilbert damping, *Phys. Rev. Lett.* **130**, 046704 (2023).

26. Bayreuther, G., Premper, J., Sperl, M., and Sander, D., Uniaxial magnetic anisotropy in Fe/GaAs (001): Role of magnetoelastic interactions, *Phys. Rev. B*. **86**, 054418 (2012).
27. Chen, L. *et al*. Robust spin-orbit torque and spin-galvanic effect at the Fe/GaAs (001) interface at room temperature. *Nature Commun.* **7**, 13802 (2016).
28. Kasai, S. *et al*. Modulation of effective damping constant using spin Hall effect, *Appl. Phys. Lett*. **104**, 092408 (2014).
29. Safranski, C., Montoya, E. A., & Krivorotov, I. N., Spin-orbit torques driven by a planar Hall current, *Nature Nanotech*. **14**, 27-30 (2019).
30. Capua, A., Rettner, C., & Parkin, S. S. P., Parametric harmonic generation as a probe of unconstrained spin magnetization precession in the shallow barrier limit, *Phys. Rev. Lett*. **116**, 047204 (2016).
31. Capua, A. *et al*. Phase-resolved detection of the spin Hall angle by optical ferromagnetic resonance in perpendicularly magnetized thin films, *Phys. Rev. B*. **95**, 064401 (2017).
32. Ou, Y., Pai, C. F., Shi, S., Ralph, D. C., and Buhrman, R. A., Origin of fieldlike spin-orbit torques in heavy metal/ferromagnet/oxide thin film heterostructures, *Phys. Rev. B* **94**, 140414(R) (2016).
33. Chen, L., Matsukura, F., & Ohno, H., Direct-current voltages in (Ga,Mn)As structures induced by ferromagnetic resonance. *Nature Commun.* **4**, 2055 (2013).
34. Gmitra, M. *et al*. Magnetic control of spin-orbit fields: A first-principle study of Fe/GaAs junctions. *Phys. Rev. Lett.* **111**, 036603 (2013).
35. Staunton, J. B. *et al*. Temperature dependence of magnetic anisotropy: An *ab initio*

approach, *Phys. Rev. B*. **74**, 144411 (2006).

36. Goennenwein, S. T. B. *et al*. Piezo-voltage control of magnetization orientation in a ferromagnetic semiconductor. *Phys. Status Solidi (RRL)* **2**, 96-98 (2008).
37. Overby, M., Chernyshov, A., Rokhinson, L. P., Liu, X., & Furdyna, J. K., GaMnAs-based hybrid multiferroic memory device, *Appl. Phys. Lett*. **92**, 192501 (2008).
38. MacNeill, D. *et al*. Control of spin-orbit torques through crystal symmetry in $WTe_2$/ferromagnet bilayers, *Nature Phys.* **13**, 300 (2017).
39. Bai, H. *et al*. Observation of spin splitting torque in a collinear antiferromagnet $RuO_2$, *Phys. Rev. Lett.* **128**, 197202 (2022).
40. Bose, A. *et al*. Tilted spin current generated by the collinear antiferromagnet ruthenium dioxide, *Nature Elect.* **5**, 267 (2022).
41. Karube, S. *et al*. Observation of spin splitter torque in collinear antiferromagnet $RuO_2$, *Phys. Rev. Lett.* **129**, 137202 (2022).
42. Šmejkal, L., Sinova, J., and Jungwirth, T., Emerging research landscape of altermagnetism, *Phys. Rev. X* **12**, 040501 (2022).
43. Hu, S. *et al*. Efficient perpendicular magnetization switching by a magnetic spin Hall effect in a noncollinear antiferromagnet, *Nature Commun.* **13**, 4447 (2022).
44. Han, R. K. *et al*. Field-free magnetization switching in CoPt induced by noncollinear antiferromagnetic $Mn_3Ga$. *Phys. Rev. B.* **107**, 134422 (2023).

## Acknowledgements

Financial support is from SFB 1277 and TRR 360 by DFG.

## Author contributions

L.C. planned the study. Y. S. and L.C. fabricated the devices, collected the data. L.C

analysed the data and did the theoretical calculations. T.N.G.M and M.K. grew the samples. S.M. and H.E. did the first principle calculations. L.C. wrote the manuscript with input from all other co-authors. All authors discussed the results and contributed to the manuscript.

## Figure captions

FIG. 1. Schematic of the microscopic mechanism of manipulation and modification of magnetism by a spin current. (**a**) The electron spins transmitted into the FM contain both transverse and longitudinal components with respect to **M**. Due to exchange coupling, the transverse component dephases and is absorbed by **M**, which gives rise to the damping-like spin-orbit torque and is responsible for changing the direction of **M**. The longitudinal component of the spin current is on average aligned with **M**, leading to additional filling of the majority band when **M** is oriented along the +**z**-direction, and an enhancement of the magnitude $M$ as well as an increase of magnetic anisotropies are expected due to the enhanced splitting of the majority- and minority-spin energy bands. (**b**) When **M** is aligned along the –**z**-direction, the spin-polarized electron enters the minority band, which can lead to a decrease of $M$ as well as a decrease of magnetic anisotropies because of the reduction of the splitting of the majority- and minority-spin energy bands. (**c**) and (**d**): The same as (**a**) and (**b**) but the polarization of the spin current is reversed, which is expected to reduce $M$ for **M**//+**z** (c) and to enhance $M$ for **M**//-**z** (**d**).

FIG. 2. (**a**) Schematic of the device for the detection of ferromagnetic resonance by time-resolved magneto-optical Kerr microscopy. (**b**) Schematic of the Pt/Al/Fe/GaAs (001) structure. (**c**) Diagram of crystallographic axes with easy and hard magnetization axes along <110> and <$\bar{1}$10> orientations. (**d**) FMR spectra for different dc currents $I$ measured at $f$ = 12 GHz and $\varphi_{I\text{-}H}$ = 90°, where $\varphi_{I\text{-}H}$ is the angle between the magnetic-field and the current direction as shown in the inset. (**e**) FMR linewidth (full width at half maximum) as a function of dc current for $\varphi_{\text{I-H}} = \pm 90^{\circ}$; solid lines are the linear fits from which the modulation amplitude $d(\Delta H)/dI$ is obtained. (**f**) $\varphi_{\text{I-H}}$-dependence of $d(\Delta H)/dI$. Error bars show the standard error of the least squares fit. The solid line is the calculation result based on the spin Hall effect of Pt when taking into account the in-plane magnetic anisotropies of Fe (Supplementary Note 4).

FIG. 3. $I$ dependence of $H_R$ measured at selected frequencies for $H$ along [110]- (**a**) and $[\bar{1}10]$-orientations (**b**) for $t_{Fe} = 2.8$ nm. For both field orientations, $H_R(-I) > H_R(+I)$ holds where $H_R(-I)$ and $H_R(+I)$ are respectively marked by red and blue arrows in each panel. (**c**) and (**d**): the same plots as (**a**) and (**b**) but for $t_{Fe} = 1.2$ nm. In (**c**) for $H$ along the [110]-orientation, $H_R(-I) > H_R(+I)$ still holds for all measured frequencies. However, for $H$ along the $[\bar{1}10]$-orientation as shown in (**d**), the relative magnitude of $H_R(-I)$ and $H_R(+I)$ depends on the excitation frequency, i.e., for $f = 12.0$ GHz, $H_R(-I) < H_R(+I)$ holds; for $f = 14.0$ GHz, $H_R(-I) \sim H_R(+I)$ holds; while for $f = 16.0$ GHz, $H_R(-I) > H_R(+I)$ holds. As shown in the upper panels, for all the devices, the charge currents are applied along the [100]-orientation, and the direction of the spin accumulation $\sigma$ is along the [010]-direction with equal projections onto the [110]- and $[\bar{1}10]$-orientations. This experimental trick allows an accurate comparison of the current induced modification for [110]- and $[\bar{1}10]$-orientations in the same device.

FIG. 4. (**a**) $f$-dependence of $dH_R/dI$ for $H$ along the easy axes ([110]- and $[\bar{1}\bar{1}0]$-orientations). (**b**) $f$-dependence of $dH_R/dI$ for $H$ along the hard axes ($[\bar{1}10]$- and $[1\bar{1}0]$-orientations). The results in (**a**) and (**b**) are obtained for $t_{Fe} = 2.8$ nm which show that $dH_R/dI$ is independent of $f$ for both easy and hard axes. (**c**) and (**d**) are the same plots as (**a**) and (**b**) but for $t_{Fe} = 1.2$ nm. The magnitude of $dH_R/dI$ along the hard axes for the thinner sample depends strongly on the excitation frequency, indicating the modification of magnetic anisotropies by spin current. The insets of (**a**)-(**d**) show the relative orientations between the current (**I** // [100], black arrows) and the magnetic-field (or magnetization), where the easy axes are represented by brown arrows and the hard axes are represented by green arrows.

FIG. 5. Summary of $t_{Fe}$ dependence of $\Delta H_A$ ($\Delta H_A = \Delta H_K$, $\Delta H_U$, $\Delta H_B$) for opposite magnetization **M** directions, where solid symbols represent **M** // +**z**-direction and open symbols represent **M** // −**z**-direction. The relative orientations between the charge current **I** and **M** are shown in the inset.

Table 1. Summary of the $\Delta H_R$-$f$ relationships induced by $\Delta H_K$, $\Delta H_U$, and $\Delta H_B$ along easy and hard axes.